\documentclass[prl,superscriptaddress,twocolumn,amsfonts,amsmath,longbibliography]{revtex4-2}
\usepackage{graphicx}
\usepackage{bm}

\begin{document}
\title{
Order in the interference of a long chain of Bose condensates with unrestricted phases
}
\author{Vasiliy Makhalov}
\affiliation{Institute of Applied Physics, Russian Academy of Sciences, Nizhniy Novgorod, Russia}
\author{Andrey Turlapov}\email{a\_turlapov@mail.ru}
\affiliation{Institute of Applied Physics, Russian Academy of Sciences, Nizhniy Novgorod, Russia}
\affiliation{N. I. Lobachevsky State University of Nizhni Novgorod, Nizhniy Novgorod, Russia}

\date{\today}

\begin{abstract}
For a long periodic chain of Bose condensates prepared in the free space, the subsequent evolution and interference dramatically depend on the difference between the phases of the adjacent and more distant condensates.
If the phases are equal, the initial periodic density distribution reappears at later times, which is known as the Talbot effect. For randomly-related phases, we have found that a spatial order also appears in the interference, while the evolution of the fringes differs with the Talbot effect qualitatively. Even a small phase disorder is sufficient for qualitatively altering the interference, though maybe at long evolution times. This effect may be used for measuring the amount of coherence between adjacent condensates and the correlation length along the chain.
\end{abstract}

\maketitle

A chain of interfering sources is a paradigm in several physics areas.
In optics, for a long chain of elements with equal phases, the initial intensity distribution reestablishes at certain propagation distances~\cite{Talbot}, which is referred to as the Talbot effect.
Similar phenomena have been observed in acoustics~\cite{AcousticTalbot1985}, vacuum electronics~\cite{DenisovTalbot1996,DenisovTalbot1999}, plasmonics~\cite{PlasmonTalbot2009}, and spintronics~\cite{SpinwaveTalbot2012}.

Solid state physics offers a variety of situations, where each chain element has either thermal or quantum fluctuations of its phase. In the Josephson-junction chains such fluctuations drive transitions between superconducting and isolating states~\cite{BradleyDoniach1984,LarkinBiasedChains1997}. Phase slips and the resulting negative interference may prevent the supercurrent from flowing through a chain~\cite{JosephsonChainPhaseSlips2010}.

Physics of ultracold atoms and molecules overlaps with condensed matter physics~\cite{BlochLowDReview2008,PitaevskiiFermiReview} and optics~\cite{AtomOpticsReview2009}. The Talbot effect has been detected~\cite{PritchardTalbot1995} for a chain of phased matter-wave sources obtained by passing a monochromatic atomic beam through a periodic grating. Fluctuations of the phases appear and may be controlled in a chain of Bose-Einstein condensates (BECs) in a one-dimensional optical lattice~\cite{DalibardBECInterference2004}.
Interference of condensates in the free space, upon lattice extinction, has been observed both for correlated and random phases~\cite{DalibardBECInterference2004,WeitzRandomBECArrayInterference2005} in the far-field diffraction regime, which is different to the near-field condition of the Talbot effect. Randomly-phased BECs surprisingly produce spatially periodic interference fringes~\cite{DalibardBECInterference2004,WeitzRandomBECArrayInterference2005}. At the qualitative level, however, the far-field interference pattern is similar to that of phase-locked BECs because the fringe period is the same~\cite{DalibardBECInterference2004}.

In this Letter, we analyze the spatial order in the long-chain or, equivalently, near-field-diffraction limit, with each source spreading over a distance much smaller than the chain length. We show that the randomly-phased elements produce a qualitatively different interference pattern in comparison to the equally-phased sources.
For equal phases, the Talbot effect is observed. For uncorrelated phases, the interference is also showing a spatial order, with period, however, different to that seen within the Talbot effect. For a partial correlation the two interference types coexist, which gives a way for measuring the amount of phase disorder and the correlation length.
Even a small disorder between distant condensates is sufficient to qualitatively change the near-field interference, but maybe for long interference times.

\begin{figure}[htb!]
\begin{center}
\includegraphics[width=\linewidth]{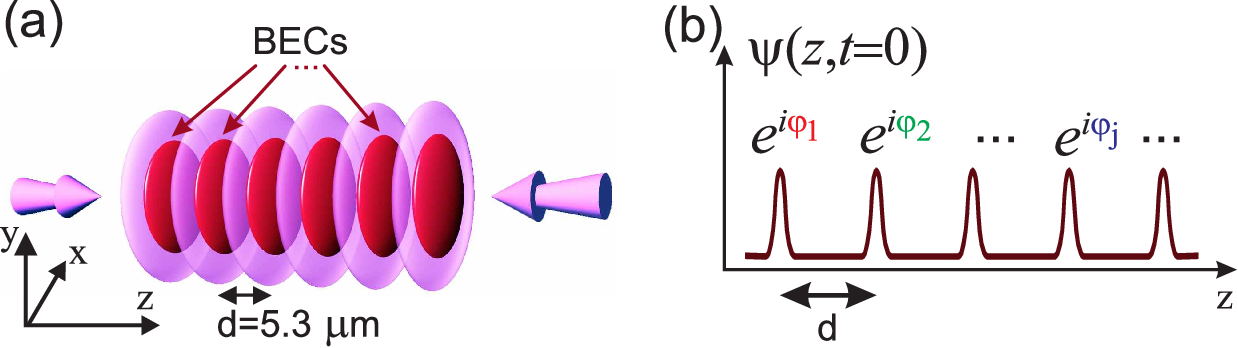}
\end{center}
\caption{(a) BECs in the lattice prior to the release and interference. The clouds of molecules shown in dark red, the standing-wave intensity shown in light purple. (b) The initial wave function along the lattice.
The density is periodic, while phase $\varphi_j$ of the $j$th condensate is generally unrestricted with respect to the phases of other BECs.}
\label{fig:InitialCondition}
\end{figure}

In the experiment, a long chain of molecular Bose-Einstein condensates interferes in the free space after preparation in and release from a one-dimensional optical lattice shown in Fig.~\ref{fig:InitialCondition}(a).
The experimental setup is similar to that of Ref.~\cite{FermiBose2DCrossover} and references therein.
The bosons are weakly-bound Li$_2$ molecules, each composed of two fermionic $^6$Li atoms.
The lattice is formed by two counter-propagating laser beams at 10.6~$\mu$m producing potential
\begin{equation}
V_s(\vec{x})=sE_{\text{rec}}\left[1-e^{-2ME_{\text{rec}}(x^2+y^2)/(\hbar\lambda)^2}\cos^2\kappa z\right],
\label{eq:OptLattice}
\end{equation}
where $\vec{x}=(x,y,z)$, $M$ is the Li$_2$ mass, $\kappa=2\pi/(10.6\text{ $\mu$m})$, $E_{\text{rec}}=\hbar^2\kappa^2/2M$ is the recoil energy, $s$ is the dimensionless lattice depth, and $\lambda=27.4$ is the anisotropy ratio of each lattice site. The lattice period is $d=5.3$ $\mu$m.
The harmonic expansion of $V_s(\vec{x})$ near each minimum gives frequencies $\omega_z\equiv2\sqrt sE_{\text{rec}}/\hbar$ and $\omega_\perp=\omega_z/\lambda$. For all data of the main text except those of Fig.~\ref{fig:SmallDisorder}, the lattice parameters are $sE_{\text{rec}}=23.3E_{\text{rec}}=165$~nK, $\omega_z/(2\pi)=1424$~Hz, and $\omega_\perp/(2\pi)=52$~Hz.
About 100 wells are populated. Central $K=50$ clouds contain nearly equal number of molecules $N$, which typically is in range 400--1100.
The molecular condensates are obtained by evaporative cooling~\cite{LeCooling,FermiBose2DCrossover} of equal mixture of atoms in the lowest-energy hyperfine states $|1\rangle$ and $|2\rangle$ \cite{HuletLiCollisions1998} at magnetic field $B=730$~G, on the Bose side of a Feshbach resonance~\cite{FeshbachReview2010}.

At the end of the preparation, time $t=0$, the optical-lattice potential (\ref{eq:OptLattice}) is quickly extinguished over $\sim1$~$\mu$s, the clouds start to spread out and interfere. The interference takes place in the time domain, without propagation of the clouds along $x$ or $y$.
The dynamics in the $z$ direction is most notable, while the expansion in the orthogonal directions is slow and unimportant here. The characteristic time scale of the interference is the Talbot time $T_d\equiv Md^2/\pi\hbar=1.69$~ms, which is set by the distance $d$ between the clouds.

\begin{figure}[htb!]
\begin{center}
\includegraphics[width=0.68\linewidth]{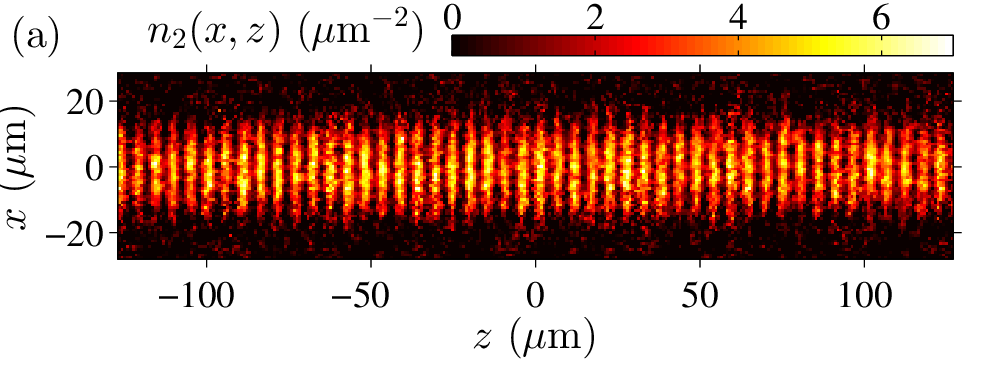}\includegraphics[width=0.3\linewidth]{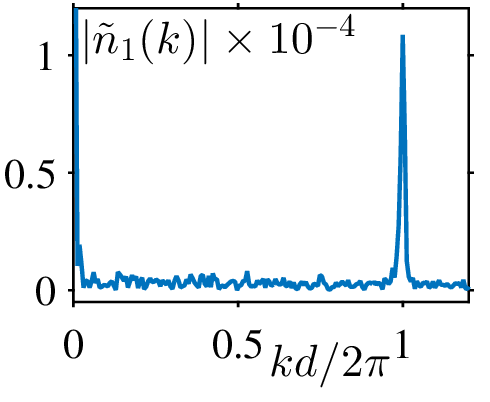}
\includegraphics[width=0.68\linewidth]{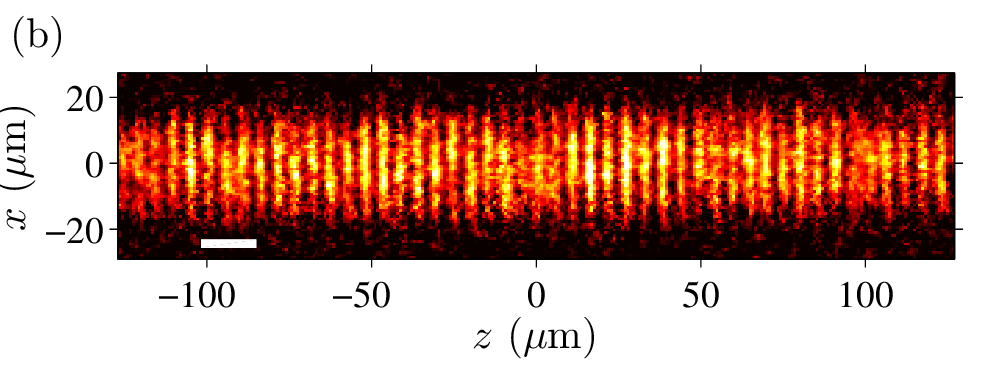}\includegraphics[width=0.3\linewidth]{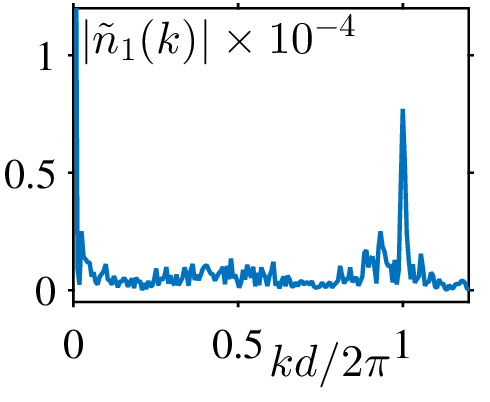}
\includegraphics[width=0.68\linewidth]{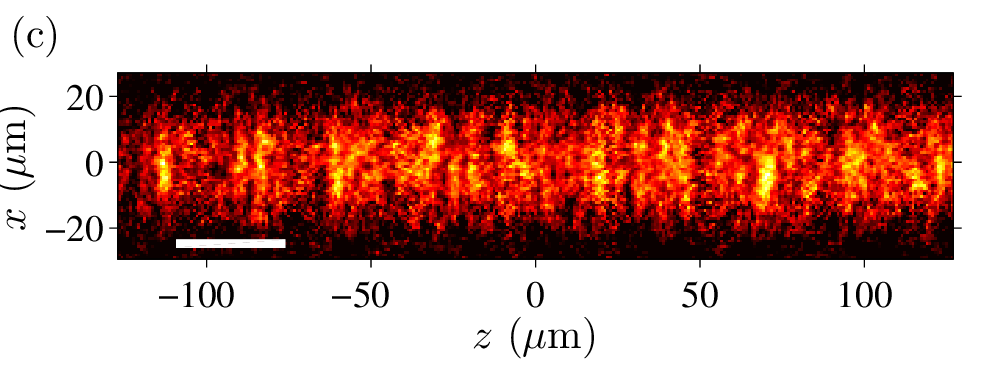}\includegraphics[width=0.3\linewidth]{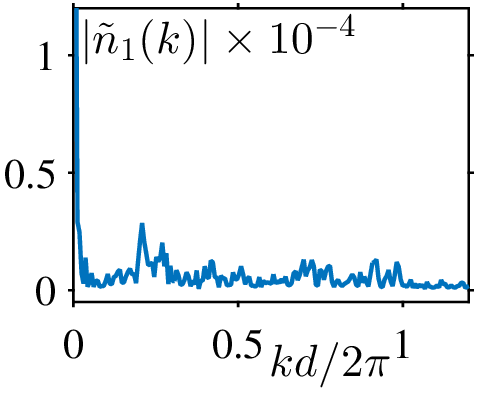}
\end{center}
\caption{Interference of a chain with nearly phased adjacent condensates: images (left) and the respective Fourier transforms $|\tilde n_1(k)|$ (right).
(a) At $t=0$, the onset of the expansion. (b) At $t=T_d$,
the initial density distribution is nearly reestablished showing the Talbot effect.
Each BEC overlaps with about 3 neighbors on the left and 3 on the right.
(c) At $t=2T_d$, the interference is governed by the random phase relation between more distant neighbors which now overlap.
In (b,c), the white bars show the full rms width along $z$ of a single condensate after the expansion.
}
\label{fig:TalbotAndCorrLength}
\end{figure}

The condensates are observed via the absorptive imaging~\cite{FermiBose2DCrossover}. A pulse of light resonant to the atomic transition is shed in the $y$ direction. The pulse duration 4 $\mu$s is shorter than any relevant scale of the matter-wave diffraction.
The shadow from the absorption is projected onto a CCD camera which allows to reconstruct the molecular density integrated along the line of sight $n_2(x,z)=\int n(\vec x)dy$.
In Fig.~\ref{fig:TalbotAndCorrLength}(a) one may see the condensates prior to the release into the free space.
The imaging destroys the state of the quantum system. For observing the interference at other times $t$, the BECs are prepared again.

The evolution of the condensates may be understood by considering the one-dimensional free-space dynamics of initial wavefunction
\begin{equation}
\psi(z,t=0)\propto\sum_{j=1}^Ke^{-(z-jd)^2/4\sigma^2}\,e^{i\varphi_j},
\label{eq:PsiInitial}
\end{equation}
where condensate half width $\sigma$ satisfies the condition of no overlapping, $\sigma\ll d$.
Such wavefunction depicted in Fig.~\ref{fig:InitialCondition}(b) has a periodic absolute value $|\psi(z-d,t=0)|=|\psi(z,t=0)|$, while phases $\varphi_j$ are generally unrestricted.
The outcome of the interference depends on the relation between phases $\varphi_j$. In the thermal equilibrium the relation between the phases settles as a result of competition: On one hand,  tunneling tends to equalize the phases, on the other hand, the interaction with the uncondensed molecules as well as quantum fluctuations randomizes the phases. The quantum Talbot effect appears in the case of equal phases $\varphi_j=\text{inv}(j)$. In the course of the evolution, the initial wave function reestablishes at times $t$ that are integer multiples of $T_d$.

Experimental demonstration of the Talbot effect can be seen in Figs.~\ref{fig:TalbotAndCorrLength}(a,b)
referring to $t=0$ and $t=T_d$ respectively. In particular, image \ref{fig:TalbotAndCorrLength}(b) (left) taken after the free evolution, at $t=T_d$, nearly reproduces the $t=0$ photo of the BECs.
In Figs.~\ref{fig:TalbotAndCorrLength}(a,b) we also show the Fourier transforms $\tilde n_1(k,t)\equiv\int n_1(z,t)e^{-ikz}dz$ of one-dimensional density distribution $n_1(z)\equiv\int n_2(x,z)dx$.
The closeness of the spectrum at $t=T_d$ to the initial spectrum is a signature of the Talbot effect.

We now check whether the BEC parameters are consistent with the phase locking required for the Talbot effect. For the data of Figs.~\ref{fig:TalbotAndCorrLength}, the evaporative cooling~\cite{FermiBose2DCrossover} finishes at lattice depth $16.6E_{\text{rec}}$. Afterwards, the lattice is slowly raised to final depth $23.3E_{\text{rec}}$.
There are $N=580\pm40$ molecules per well, where the error is the standard deviation found from several experimental repetitions.
Prior to the release, the condensates occupy the lowest Bloch band, which is seen from the chemical-potential value
$\mu=\hbar\omega_\perp\sqrt{2N(a_{\text{mol}}/l_z)\sqrt{2/\pi}}=0.36\hbar\omega_z<\hbar\omega_z$, where $l_z\equiv\sqrt{\hbar/(M\omega_z)}$ and $a_{\text{mol}}=0.6\,a=1520$ Bohr is the molecule-molecule \textit{s}-wave scattering length expressed via the atom-atom scattering length $a$~\cite{Petrov}.
The temperature is estimated from images at $t=0$ by fitting the data with the bimodal distribution typical for a 2D BEC~\cite{FermiBose2DCrossover}, which gives average temperature $T=0.45T_{\text{BEC 2D}}$, where
$T_{\text{BEC 2D}}=\hbar\omega_\perp\sqrt{6N}/\pi=0.69\hbar\omega_z$ is the condensation temperature of a 2D noninteracting Bose gas in a harmonic trap. Such fit, however, may overestimate the temperature~\cite{DalibardBECandBKT2008}.
About $1.5\%$ of the molecules are thermally excited to the 1st Bloch band. The tunnel time for a single molecule in the lowest band is 190 ms. The Bose statistics, however, should enhance the tunnel rate by a factor $N(1-T^2/T_{\text{BEC 2D}}^2)$, which gives the tunnel time $\tau_{\text{tun}}=410$~$\mu$s. The thermal dephasing time may be estimated as $\hbar/T=360$~$\mu$s. The tunneling and dephasing rates are about the same, which, therefore, leaves the possibility of phase locking.
The interaction-induced phase mismatch between the adjacent condensates is small for all reported experiments and may be estimated as $\delta\varphi=(\tau_{\text{tun}}E_{\text{c}}/(4\hbar))^{1/4}\sim0.1\ll2\pi$, where $E_{\text{c}}=2\,d\mu/dN$ \cite{BoseChainFluctPitaevskii2001}.

\begin{figure}[htb!]
\begin{center}
\includegraphics[width=0.68\linewidth]{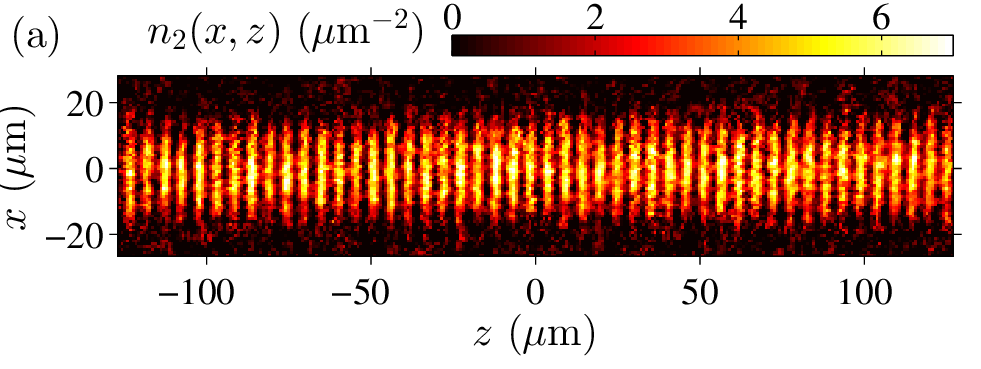}
\includegraphics[width=0.3\linewidth]{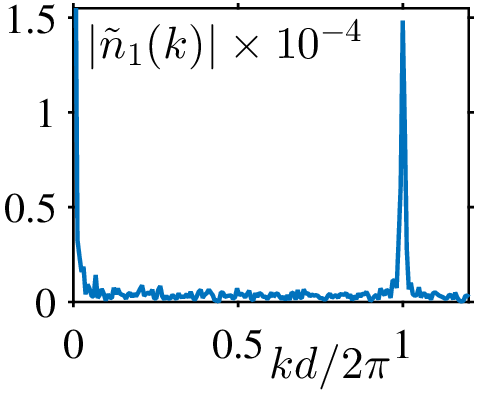}
\includegraphics[width=0.68\linewidth]{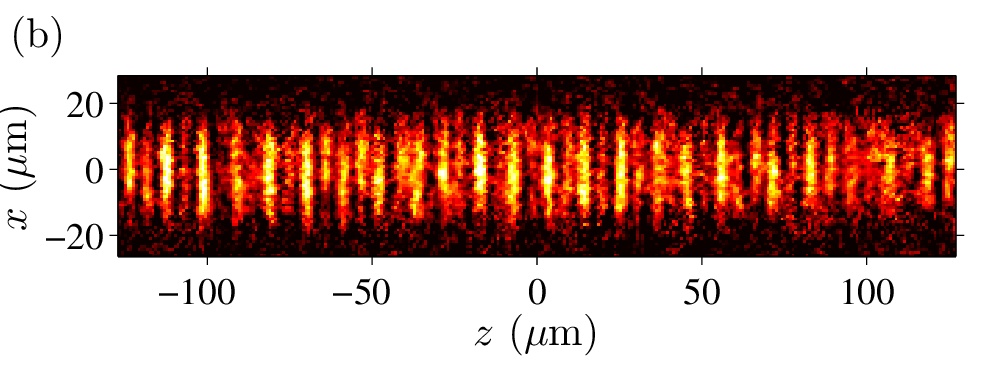}
\includegraphics[width=0.3\linewidth]{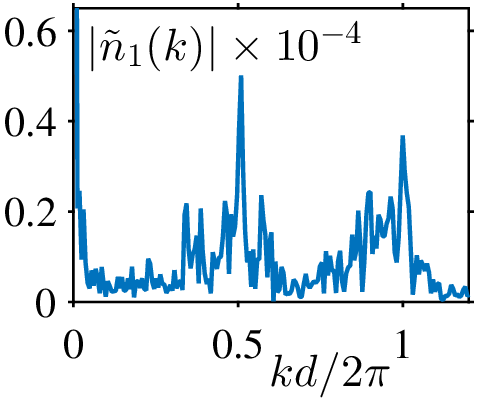}
\includegraphics[width=0.68\linewidth]{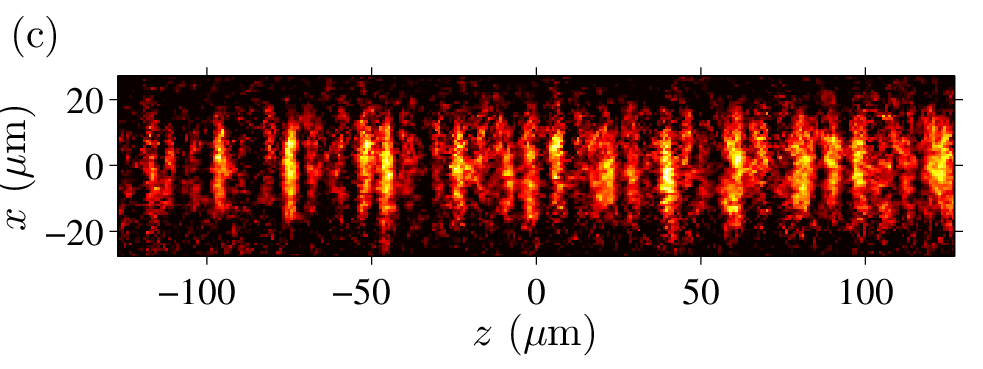}
\includegraphics[width=0.3\linewidth]{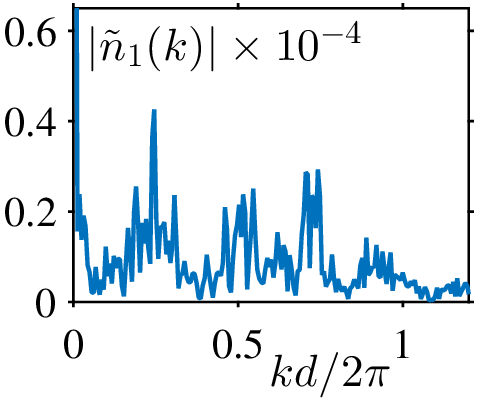}
\end{center}
\caption{Interference of a chain where the adjacent condensates have nearly random relative phases: images (left) and the respective Fourier transforms $|\tilde n_1(k)|$ (right).
(a) At $t=0$, the onset of the expansion. (b) At $t=T_d$,
the principal harmonic in the density distribution is at about $k=\pi/d$ corresponding to period $2d$.
(c) At $t=2T_d$, the principal harmonic corresponds to period $\simeq4d$.
}
\label{fig:Uncorrelated}
\end{figure}

An evolution, which qualitatively differs from the Talbot effect, is seen for condensates which undergo less cooling. Now the evaporative cooling~\cite{FermiBose2DCrossover} proceeds only down to lattice depth $23.3E_{\text{rec}}$. Fig.~\ref{fig:Uncorrelated}(a) shows the image of the condensates immediately prior to the release, while Figs.~\ref{fig:Uncorrelated}(b,c) display the interference at $t=T_d$ and $t=2T_d$ respectively.
Both at $t=T_d$ and $t=2T_d$ there is a spatial order in the density, which may be seen from the respective spatial spectra $\tilde n_1(k,t)$. The spatial period in both cases, however, is larger than in the initial density distribution. For $t=T_d$ the Fourier transform has the lowest-momentum peak at about $k=\pi/d$, which corresponds to period $2d$. For $t=2T_d$ the principal harmonic is near $k=\pi/(2d)$ corresponding to period $4d$.
The fringe straightness along $x$ also confirms that the molecules are condensed in each well.

Estimating the BEC parameters for Figs.~\ref{fig:Uncorrelated}, we find $T=0.62T_{\text{BEC 2D}}$ and $N=440\pm20$. This gives thermal dephasing time $\hbar/T=300$~$\mu$s and Bose-enhanced tunnel time $\tau_{\text{tun}}=710$~$\mu$s, making the system more prone to dephasing than in the case where the Talbot effect is observed in Figs.~\ref{fig:TalbotAndCorrLength}.

\begin{figure}[htb!]
\begin{center}
\includegraphics[width=0.6\linewidth]{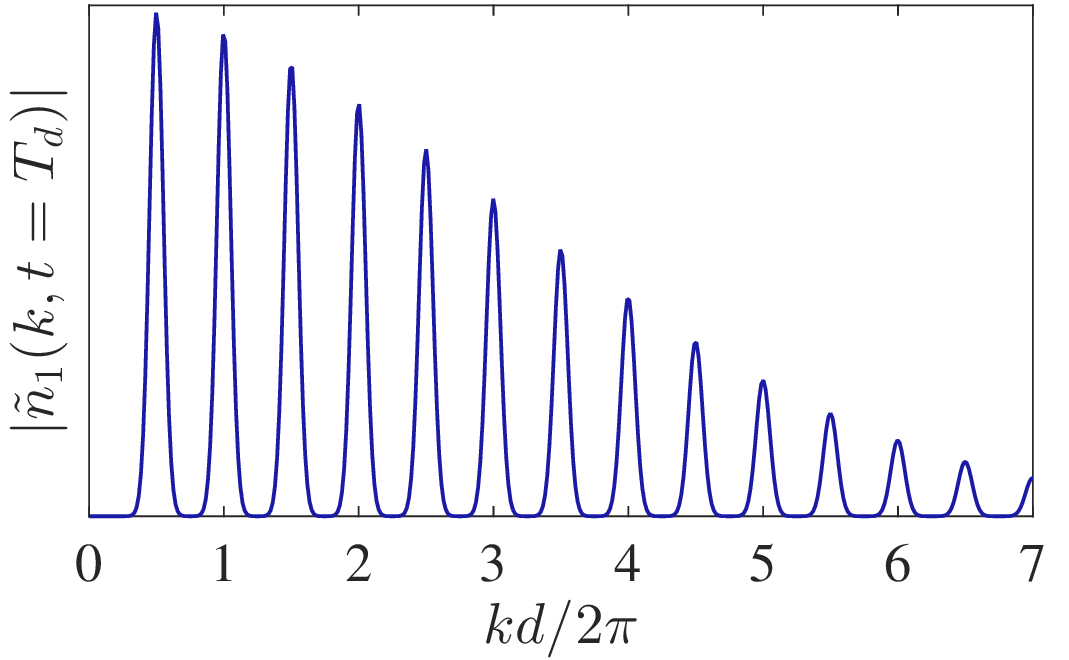}
\end{center}
\caption{The absolute value of density spectrum~(\ref{eq:TalbotDensitySpectrum}) at $t=T_d$. The principal harmonic lies at $k=\pi/d$ corresponding to spatial period $2d$.}
\label{fig:FresnelSpectrum}
\end{figure}

\begin{figure}[htb!]
\begin{center}
\includegraphics[width=0.68\linewidth]{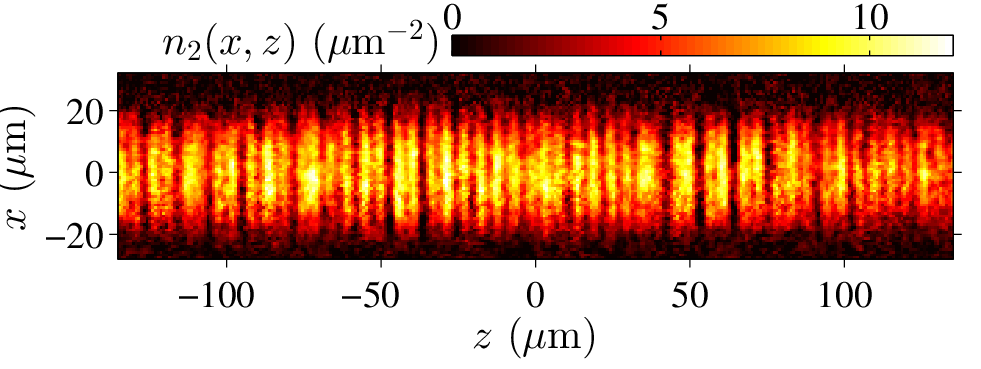}
\includegraphics[width=0.3\linewidth]{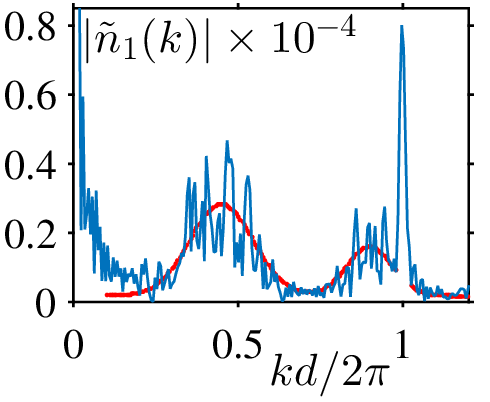}
\end{center}
\caption{Interference at $t=T_d$ of a chain with a partial phase disagreement between the adjacent condensates: image $n_2(x,z)$ (left) and the respective Fourier transform $|\tilde n_1(k)|$ (right). Fit for finding the centers of the phase-randomization-related peaks is shown as the red curve.
}
\label{fig:SmallDisorder}
\end{figure}

We interpret Figs.~\ref{fig:Uncorrelated} as the near-field interference of molecular BECs whose phases $\varphi_j$ are random relative to each other. A model qualitatively explaining the observation is presented below. At $t=0$, the wavefunction is taken in form (\ref{eq:PsiInitial})
with phases $\varphi_j$ randomly distributed over interval $[0,2\pi]$. For later times $t$, Fourier transform $\tilde{n}_1(k,t)=\int|\psi(z,t)|^2e^{-ikz}dz$ may be calculated.
In the long-chain limit $K\rightarrow\infty$, $\tilde{n}_1(k,t)$ takes form~\cite{MarTurUnpub}
\begin{align}
\tilde{n}_1(k,t)&\propto\frac{2\pi}d\delta(k) + \frac{\sqrt{\pi K}}2 e^{-k^2\sigma^2/2}\times\nonumber\\ &\sum_{j=-\infty,\,j\neq0}^{\infty} e^{-(j-kdt/T_d\pi)^2 d^2/8\sigma^2} e^{i\varphi_j'},
\label{eq:TalbotDensitySpectrum}
\end{align}
where $\varphi_j'$ are random phases satisfying condition $\varphi_j'=-\varphi_{-j}'$. Due to condition $\sigma\ll d$, the gaussians in sum~(\ref{eq:TalbotDensitySpectrum}) are narrow peaks. Therefore, at time $t$ the density is a sum of harmonics with wave vectors $k=j\pi T_d/(td)$ for integer $j$. This means that the density distribution $n_1(z,t)$ is periodic in space with spatial period
\begin{equation}
\frac{2\pi}k=d\frac{2t}{T_d}.\label{eq:RandomPhasePeriod}
\end{equation}
Therefore, there is a spatial order in the density, in agreement with the data of Figs.~\ref{fig:Uncorrelated}(b,c). The calculated density spectrum is shown in Fig.~\ref{fig:FresnelSpectrum}.
The spectrum envelope is preserved during the evolution, while the peaks come closer and become narrower. The density-spectrum formula~(\ref{eq:TalbotDensitySpectrum}) is valid for $t\gg T_d\sigma/d$, which is not a stringent constraint because $\sigma\ll d$. This implies that the analysis is correct for $t=T_d$ as well as for even earlier times.

Whenever the phases are not completely random, the Talbot effect coexists with the random-phase interference. A combination of these two effects is seen in Fig.~\ref{fig:SmallDisorder}.
In the Fourier transform, the tall narrow peak at $k=2\pi/d$ is due to the correlations between the condensates. There is also a wider peak at $k=0.45\times2\pi/d$ and its 2nd harmonic at $k=0.9\times2\pi/d$. Their centers are found by fitting a curve [Fig.~\ref{fig:SmallDisorder}(right)], which is constrained to have maxima at a principle wave vector and a twice larger value. These two peaks are the signature of a partial uncorrelation in the phases, because no peak below $k=2\pi/d$ may appear within the Talbot effect.
The peak centers are somewhat below values $k=0.5\times2\pi/(2d)$ and $k=2\pi/(2d)$ respectively predicted within model (\ref{eq:TalbotDensitySpectrum}) of fully uncorrelated BECs. This small leftward shift is due to the mean-field repulsion as we have found from a numerical simulation based on the Gross--Pitaevskii equation \cite{PitaevskiiGPE1961RusEng,GrossGPE1961}. The repulsion effectively speeds up the time in Eqs.~(\ref{eq:TalbotDensitySpectrum}) and (\ref{eq:RandomPhasePeriod}) shifting the peaks to lower momenta.
The leftward shift is not due to the finite chain length because this would have shifted proportionally both the peaks related to the phased and unphased interference.

A crossover from the Talbot effect to the random-phase interference is observed by varying lattice depth $s$ \cite{SM2019}.

The relation between the peaks originating from the two interference scenarios, as in Fig.~\ref{fig:SmallDisorder}, may be used for measuring the amount of correlation between adjacent condensates. The correlation measurement can be done in a single experimental repetition. In principle, even a small departure from full correlation may be detected because any randomness changes the interference qualitatively. For example, in the spectrum of Fig.~\ref{fig:TalbotAndCorrLength}(b), the little bulging near $k=0.5\times2\pi/(2d)$ is a signature of a small decoherence between the condensates.

A seeming controversy between the interference in Fig.~\ref{fig:TalbotAndCorrLength}(b) and Fig.~\ref{fig:SmallDisorder} may be noted by analyzing the respective parameters.
For the experiment of Fig.~\ref{fig:SmallDisorder}, the BEC chain is evaporatively cooled down to lattice depth $22.8E_{\text{rec}}$, and the condensates are released at this depth. For each cell $N=1060\pm80$ and $T=0.40T_{\text{BEC 2D}}$. The resulting ratio of Bose-enhanced tunnel time $\tau_{\text{tun}}=200$~$\mu$s and thermal dephasing time $\hbar/T=310$~$\mu$s nominally makes this chain more prone to phase locking than in the case of Fig.~\ref{fig:TalbotAndCorrLength}. Nevertheless in Fig.~\ref{fig:TalbotAndCorrLength}(b), the Talbot effect is more pronounced.
This controversy may be resolved by accounting for a stray magnetic-field curvature, which acts on atoms and, together with lattice~(\ref{eq:OptLattice}), produces potential
\begin{equation}V(\vec x)=V_s(\vec x)+M\omega_B^2\left(\frac{z^2}2+\frac{x^2}2-y^2\right).\end{equation}
The primary curvature source is the electromagnet coils for tuning the interactions near the Feshbach resonance. By another pair of coils, the curvature is canceled down to $\omega_B^2/(2\pi)^2=0_{-1.6}^{+7}$~Hz$^2$ \cite{SM2019}. The stray curvature, which remains within the error margins, causes small spatial variation of the cell depths. Evaporative cooling creates small unevenness in the chemical potentials which in turn brings about Josephson oscillations and pseudo randomization of the BEC phases. Such pseudo randomization may be the reason for stronger randomness-related peaks in Fig.~\ref{fig:SmallDisorder} (right).

In the light of the two interference types clearly combined in Fig.~\ref{fig:SmallDisorder}, the spectrum of Fig.~\ref{fig:Uncorrelated}(b) may be reanalyzed: It is possible that the spike at sharply $k=2\pi/(2d)$ is due to a small residual coherence.

Infinitely-extended spatial correlations are not possible since the chain is one-dimensional~\cite{BradleyDoniach1984}. The phase correlation $\langle\cos(\varphi_i-\varphi_j)\rangle$ should decay either as a power law $\propto|i-j|^{-\nu}$ or exponentially $\propto\alpha^{-|i-j|}$ \cite{BradleyDoniach1984,BoseChainFluctPitaevskii2001}. The decay of correlations may be observed in the near-field interference by varying the number of neighbors, each BEC interferes with. This number depends on the free-evolution time $t$. In Figs.~\ref{fig:TalbotAndCorrLength}(b,c), one may see that the interference changes qualitatively with the increasing number of the involved neighbors.
At $t=T_d$ the fringes have period $d$ [Fig.~\ref{fig:TalbotAndCorrLength}(b)] consistently with the Talbot effect, which means that the interfering clouds are phased with each other.
The experiment is repeated with longer evolution time $t=2T_d$ [Fig.~\ref{fig:TalbotAndCorrLength}(c)].
The Fourier transform has the strongest peak near $k=\pi/(2d)$, in agreement with prediction (\ref{eq:RandomPhasePeriod}) for the random-phase interference. This means that the distant clouds mostly contributing to the interference are uncorrelated.

The correlation length may be estimated from the data of Figs.~\ref{fig:TalbotAndCorrLength}(b,c). Taking that the initial molecular wavefunction is a gaussian with rms half width $\sqrt{\hbar/(2M\omega_z)}$ and assuming its free expansion, one finds that by $t=T_d$ each cloud spreads to the rms half width of 8~$\mu$m, while at $t=2T_d$ the half width is 16~$\mu$m. Therefore, the correlation length is 15--20~$\mu$m. The precision may be improved with an advent of a more quantitative model. A much more distant breakdown of correlation may be measured in principle, though this would require a large interference time $t$.

In conclusion, in an infinite chain of Bose condensates with random phases, a spatial order forms as a result of interference. The spatial period appears shortly after the onset of the expansion and grows linearly with time. For partially correlated phases such interference combine with the Talbot effect. The interplay between these two types of the interference may be used for measuring the phase difference between the adjacent condensates as well as for measuring the correlation length.
These effects may be seen in lattices with other periods $d$ since $d$ is not limited to any particular range both in the quantum Talbot effect and the random-phase-interference model of Eqs. (\ref{eq:PsiInitial})--(\ref{eq:RandomPhasePeriod}).

\begin{acknowledgments}
The authors acknowledge K.~A.~Martiyanov's participation in the early stages of this work. V.~M. has been supported by Russian Foundation for Basic Research grant 15-02-08464 and by the Presidium of the Russian Academy of Sciences (Programme 1.4 ``Actual problems of low temperature physics''), A.~T. has been supported by Russian Science Foundation project 18-12-00002.
\end{acknowledgments}


%

\clearpage
\appendix
\onecolumngrid 
\section{Supplemental Material: ``Order in the interference of a long chain of Bose condensates with unrestricted phases'' by V.~Makhalov and A.~Turlapov}

\begin{figure}[htb!]
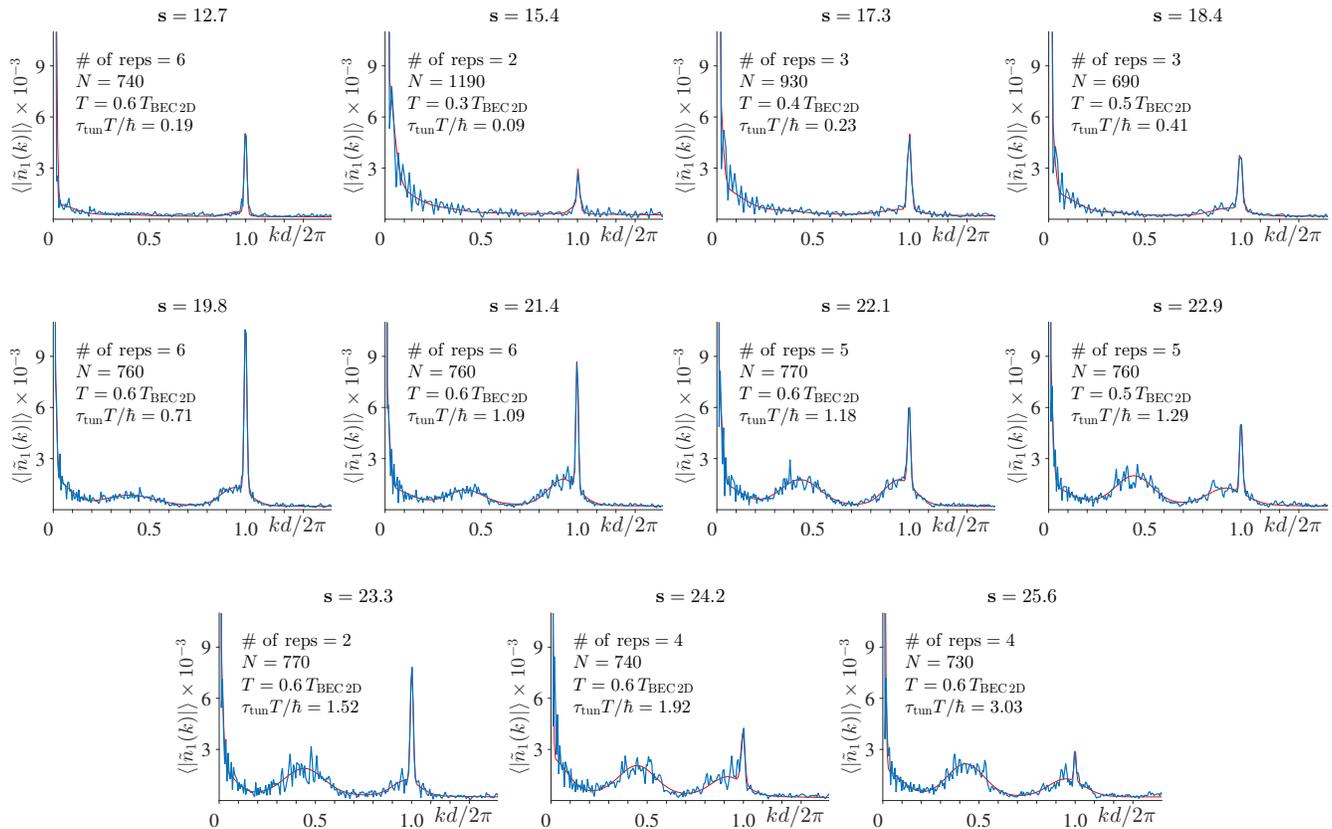
\begin{center}

\includegraphics[width=0.24\linewidth]{Fig6sm_2018-08-12-fit_mean_spectr_all_free_Cooling_Depth_10000.eps}
\includegraphics[width=0.24\linewidth]{Fig6sm_2018-08-12-fit_mean_spectr_all_free_Cooling_Depth_8000.eps}
\includegraphics[width=0.24\linewidth]{Fig6sm_2018-08-12-fit_mean_spectr_all_free_Cooling_Depth_7000.eps}
\includegraphics[width=0.24\linewidth]{Fig6sm_2018-08-12-fit_mean_spectr_all_free_Cooling_Depth_6500.eps}
\\\vspace{6mm}
\includegraphics[width=0.24\linewidth]{Fig6sm_2018-08-12-fit_mean_spectr_all_free_Cooling_Depth_6000.eps}
\includegraphics[width=0.24\linewidth]{Fig6sm_2018-08-12-fit_mean_spectr_all_free_Cooling_Depth_5500.eps}
\includegraphics[width=0.24\linewidth]{Fig6sm_2018-08-12-fit_mean_spectr_all_free_Cooling_Depth_5300.eps}
\includegraphics[width=0.24\linewidth]{Fig6sm_2018-08-12-fit_mean_spectr_all_free_Cooling_Depth_5100.eps}
\\\vspace{6mm}
\includegraphics[width=0.24\linewidth]{Fig6sm_2018-08-12-fit_mean_spectr_all_free_Cooling_Depth_5000.eps}
\includegraphics[width=0.24\linewidth]{Fig6sm_2018-08-12-fit_mean_spectr_all_free_Cooling_Depth_4800.eps}
\includegraphics[width=0.24\linewidth]{Fig6sm_2018-08-12-fit_mean_spectr_all_free_Cooling_Depth_4500.eps}
\end{center}
\caption{The spatial spectra $\langle|\tilde n_1(k)|\rangle$ for cooling down to lattice depth $s$, hold, release, and evolution for time $t=T_d$. The spectra are averaged over a number of repetitions, which is noted on each panel, together with depth $s$, average $N$, $T$, and $\tau_{\text{tun}}T/\hbar$, where the latter is the ratio of the collective-tunneling time $\tau_{\text{tun}}$ to the thermal dephasing time $\hbar/T$. The solid curves are fits applied to $\langle|\tilde n_1(k)|\rangle$.
}\label{fig:CoherenceSpectra}\end{figure}
\twocolumngrid

\subsection{Crossover from coherent to incoherent BECs}

The density spectra $\tilde n_1(k)$ at $t=T_d$, such as shown in Fig.~\ref{fig:CoherenceSpectra}, carry the information about the coherence between the condensates.
The coherence may be judged from the relative strengths of the features related to the two interference scenarios. The sharp peak at $k=2\pi/d$ originates from the Talbot effect due to some correlation of the BEC phases $\varphi_j$. The wider peaks around $k=\pi/d$ and $k=2\pi/d$ are due to the partial phase uncorrelation.

In Fig.~\ref{fig:CoherenceSpectra}, a crossover from the Talbot-effect-dominated interference to the primarily uncorrelated-BEC interference is evident, with the spatial spectra presented in the order of decreasing correlation. Experimentally, the correlation is controlled by the lattice depth $s$, noted in Fig.~\ref{fig:CoherenceSpectra} at the top of each panel. For
all the measurements the gas is evaporatively cooled to a final lattice depth $s$, then held at this depth for 1~s, released at $t=0$, and imaged at $t=T_d$. The depth $s$ determines the 1-particle tunnel time.
Each spectrum $\langle|\tilde n_1(k)|\rangle$ of Fig.~\ref{fig:CoherenceSpectra} is the result of averaging over 2--6 repetitions. This is different from the data of Figs.~2, 3, and 5, where single measurements $|\tilde n_1(k)|$ are used, without averaging. In Figs.~\ref{fig:CoherenceSpectra} the increase of $s$ brings about the reduction of the narrow peak at $k=2\pi/d$.

For mostly coherent chains in Fig.~\ref{fig:CoherenceSpectra}, the relative sizes of the decoherence-related peaks are in contradiction with the simple model of Eqs.~(2)--(3). This model predicts that the peak at $k=\pi/d$ is taller than the one at $k=2\pi/d$, as seen in Fig~4. Contrary to this prediction, for $s=17.3$--$21.4$, the peak at $k=\pi/d$ is smaller. For $s=22.1$ the two peaks are of the same height, while for even deeper lattices the peak sizes are in qualitative agreement with Eqs.~(2)--(3).

\begin{figure}[htb!]\begin{center}
\includegraphics[width=\linewidth]{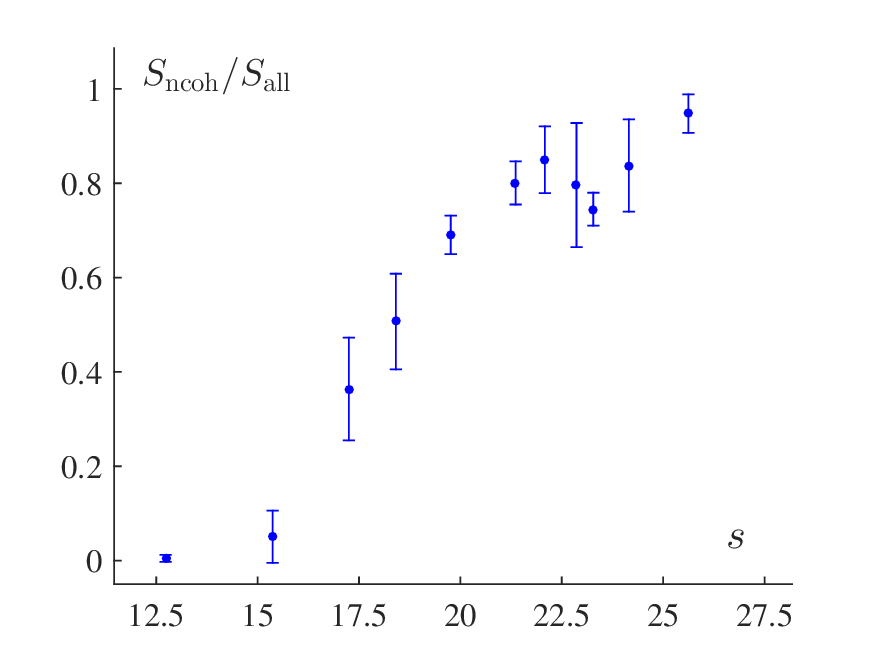}
\end{center}
\caption{The incoherent contribution to the interference spectrum vs the lattice depth $s$. Each datum corresponds to a panel in Fig.~\ref{fig:CoherenceSpectra} labeled with the respective $s$ value.}\label{fig:CoherenceVsS}
\end{figure}
\begin{figure}[htb!]\begin{center}
\includegraphics[width=\linewidth]{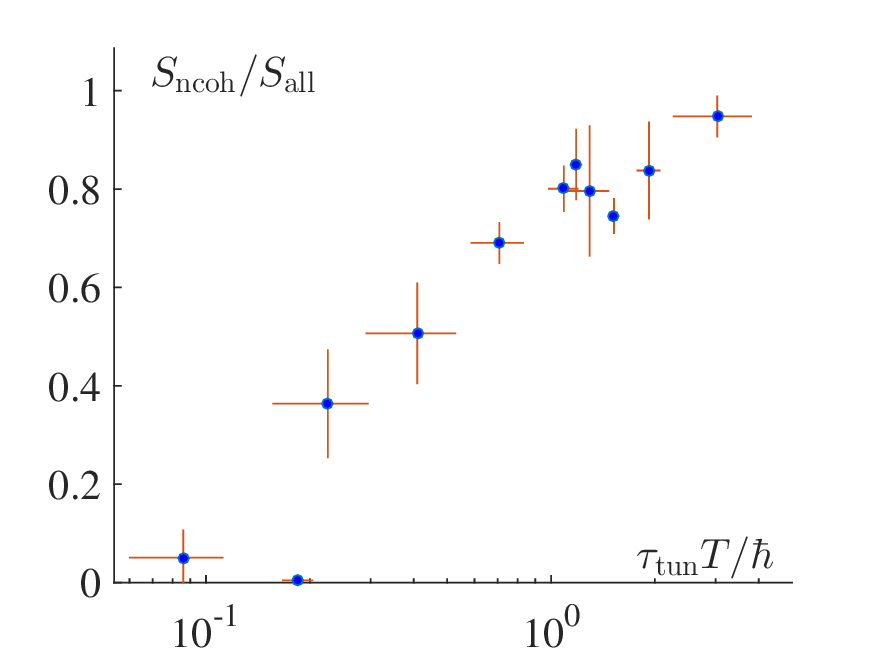}\end{center}
\caption{The incoherent contribution to the interference spectrum vs the ratio of the collective tunneling time $\tau_{\text{tun}}$ to the thermal dephasing time $\hbar/T$.}\label{fig:CoherenceVsDephasing}
\end{figure}

The coherence between the BECs is quantified by relating the area under the wide peaks in $|\tilde n_1(k)|$, $S_{\text{ncoh}}$, to the area under all 3 peaks, $S_{\text{all}}$. The ratio $S_{\text{ncoh}}/S_{\text{all}}$ vs $s$ is shown in Fig.~\ref{fig:CoherenceVsS}.
For obtaining $S_{\text{ncoh}}$ and $S_{\text{all}}$, we fit each unaveraged spectrum $|\tilde n_1(k)|$ with 6 gaussians: 1 gaussian for the narrow Talbot-interference peak at $k=2\pi/d$; 2 more gaussians for the wider peaks near $k=\pi/d$ and $k=2\pi/d$; and 3 gaussians centered at $k=0$ for the low-momentum spectrum.
The value $S_{\text{all}}$ is calculated as the area under 3 gaussians: the sharp Tablot-related gaussian at $k=2\pi/d$ and the two wider gaussians near $k=\pi/d$ and $k=2\pi/d$. The value $S_{\text{ncoh}}$ equals $S_{\text{all}}$ less the area under the sharp Tablot-related gaussian at $k=2\pi/d$. The ratios $S_{\text{ncoh}}/S_{\text{all}}$ are calculated for each experimental repetition, and the values for the same $s$ are averaged and reported in Fig.~\ref{fig:CoherenceVsS}. The error bars represent $\pm1$ standard deviation.

As a control parameter for the coherence of the chain elements, one may use $\tau_{\text{tun}}T/\hbar$, which is the ratio of the collective tunneling time $\tau_{\text{tun}}$ to the thermal dephasing time $\hbar/T$. The advantage of this parameter over $s$ is the disconnection from the details of the preparation. In Fig.~\ref{fig:CoherenceVsDephasing}, the same data as in Fig.~\ref{fig:CoherenceVsS} is reported for $\tau_{\text{tun}}T/\hbar$ as the control parameter.
The error bars of $\tau_{\text{tun}}T/\hbar$ are $\pm1$ standard deviation.

\subsection{Magnetic-field curvature cancelation}

The curvature of the magnetic field originating from the Feshbach coils is canceled down to nominally $\omega_B=0$ by another pair of coils. The cancelation current is tuned for maximizing the Talbot-type interference, such as in Fig.~\ref{fig:TalbotAndCorrLength}(b). By deliberately applying a small curvature, nominally either $\omega_B^2/(2\pi)^2=-3.2$~Hz$^2$ or $\omega_B^2/(2\pi)^2=14$~Hz$^2$, we observe a degradation of the Talbot interference. This suggests that the error margins are the halves of these curvature values, i.~e. the curvature is canceled down to $\omega_B^2/(2\pi)^2=0_{-1.6}^{+7}$~Hz$^2$.

Without the cancelation, at $\omega_B/(2\pi)=12$~Hz, we are unable to see the Talbot effect even for deepest cooling. Only the random-phase-type interference, like that of Figs.~\ref{fig:Uncorrelated}, is detected.

Also the lattice beams may have a small power imbalance which creates curvature $\lesssim0.5$~Hz$^2$ along the $z$ direction. We neglect this curvature in comparison to the error margins of $\omega_B$.

\end{document}